\documentclass[aip,apl,preprint,nofootinbib]{revtex4-2}
\usepackage{amsmath,amssymb,graphicx,xcolor}
\usepackage{booktabs}

\begin{document}
\title{Acoustic Resonance Distribution for Core–Shell Scatterers}

\author{Naruna E. Rodrigues }
\email{Corresponding author: naruna@usp.br}
\affiliation{Instituto de Física de São Carlos, Universidade de São Paulo, São Carlos 13566-590, Brazil}

\author{Gilberto Nakamura}
\affiliation{FACOM, Universidade Federal de Uberlândia, Uberlândia 38400-902, Brazil}

\author{Odemir M. Bruno}
\affiliation{Instituto de Física de São Carlos, Universidade de São Paulo, São Carlos 13566-590, Brazil}

\author{Alexandre S. Martinez}
\affiliation{Faculdade de Filosofia, Ciências e Letras de Ribeirão Preto, Universidade de São Paulo, Ribeirão Preto 14040-900, Brazil}

\begin{abstract}
Acoustic metamaterials can exhibit unusual effective properties through mechanisms such as energy localization and resonant behavior in their constituent building blocks. Here, we investigate the internal acoustic energy of fluid core–shell spheres and map how resonances emerge from the interplay between material contrasts and shell geometry. The resulting phase diagram reveals an organized resonant landscape across distinct impedance-contrast regimes. From this structure, we obtain two compact predictive relations: one for resonance existence and another for the spectral recurrence of successive peaks. Together, they enable the identification of material and geometrical combinations associated with targeted resonant responses. The framework is further tested against representative systems from the literature, reproducing their observed behavior. These results provide a physics-based route for screening and tailoring resonant core–shell building blocks before full-wave numerical simulations.

\end{abstract}
\maketitle

\section{Introduction}
\label{sec:introduction}

The concentration and storage of acoustic energy in subwavelength regions is one of the main problems in the design of acoustic metamaterials and structured media. In this context, locally resonant sonic crystals, Helmholtz-resonator arrays, and space-coiling structures with large impedance contrasts and internal resonances can enhance energy storage, usually associated with high-quality-factor modes and enhanced local fields~\cite{liu2000,fang2006ultrasonic,ma2014acoustic,Cummer2016}. This effect is important for several applications such as acoustic energy harvesting~\cite{qi2020acoustic,wang2021recent}, subwavelength imaging, and sensing~\cite{lemoult2011resonant,zhu2016flat}. Therefore, understanding how the material parameters, the geometry, and the impedance mismatch at the interfaces control the internal energy distribution is essential for designing acoustic metamaterials.

More recently, internal-energy studies show that a strong impedance mismatch in a fluid sphere can generate monopolar and higher-order resonances, and the internal field is better suited than far-field scattering for characterizing energy localization in inhomogeneous structures~\cite{Arruda:10,Arruda:11,Arruda_2012,PhysRevA.87.043841,Arruda:17,doi:10.1080/17455030.2020.1738590,chew2019acoustic,rodrigues2026}. For layered spheres, the fluid-sphere scattering problem has been solved via partial-wave expansions with continuity of pressure and radial velocity at both interfaces~\cite{mcnew2009}, and the fluid core can strongly influence the shell's low-frequency monopole response~\cite{marston2025}. Core--shell and fluid-like particles have also been cited in contexts such as virus-like particles, biomedical applications, and subwavelength liquid--liquid lenses~\cite{gordon2020sars,lamb2024mutation,kumari2022critical,boby2023sars,sanderson2023molnupiravir,Veras2022.11.21.517338,ZininAllenLevin2005PRE,perezlopez2019}, as well as coated particles for scattering cancellation~\cite{leaoneto2016}; however, these cases are outside the frequency range considered here and are included only as background. Overall, the resonant response of core--shell carriers affects energy storage, scattering, and acoustic forces, indicating that the internal distribution of material properties can be used to modify them. The open question remains: under which conditions do different resonant regimes appear, and how are these regimes controlled by the core--shell parameters?

Although several advances have been made, most previous studies on core--shell acoustic structures have focused on specific configurations or relatively narrow parameter ranges, including scattering cancellation~\cite{leaoneto2016}, subwavelength focusing~\cite{perezlopez2019}, and the monopole resonance of bubble shells~\cite{marston2025}. A systematic description of the resonant response across the broader material and geometric parameter space is therefore still lacking. In particular, it remains important to understand how the impedance contrasts between the core, shell, and surrounding medium combine with the shell thickness to control the internal energy distribution and scattering response.

Here, this problem is addressed through a resonance phase diagram that organizes the parameter space of a fluid core--shell sphere according to the occurrence and spectral distribution of resonances. The diagram reveals how impedance ordering separates resonant from non-resonant regions and how variations in shell thickness shift the resonance positions. This description is further condensed into two closed-form relations. The first determines, from the impedance contrasts and shell thickness alone, whether a given configuration is expected to resonate, providing a rapid pre-screening of candidate material combinations before the full spectrum is calculated. The second establishes a recurrence relation between successive resonances associated with distinct multipolar modes, allowing the position of the next resonance to be estimated from the preceding one together with the material contrasts and shell thickness.

In this way, the core--shell sphere provides a minimal model for understanding how material contrast and internal layering can be used to tune energy localization, scattering, and resonant amplification in structured acoustic media. At the same time, the model is restricted to longitudinal acoustic waves in fluid media. Therefore, transverse modes, shear-wave conversion, elastic deformations of solid layers, and thermal effects are not included. These assumptions make it possible to isolate the role of impedance contrast and shell geometry in the resonant response, but they also define the range of validity of the present formulation.

This article is organized as follows. In Sec.~\ref{sec:scatp} presents the governing equations for a fluid core–shell sphere. Sec.~\ref{sec:results} applies partial-wave expansions and boundary conditions at both interfaces to obtain the scattering and internal coefficients. Sec.~IV derives separate expressions for the acoustic energy stored in the core and shell. Sec.~\ref{sec:conclusions} discusses the main implications and outlook. The appendix then works through the impedance-contrast regimes in detail, explicitly connecting them to the phase diagram and to the closed-form classifier and recurrence relations.

    \section{Core--shell acoustic scattering model}\label{sec:scatp}
    
\begin{figure}[!htbp]
  \centering
  \includegraphics[width=1\linewidth]{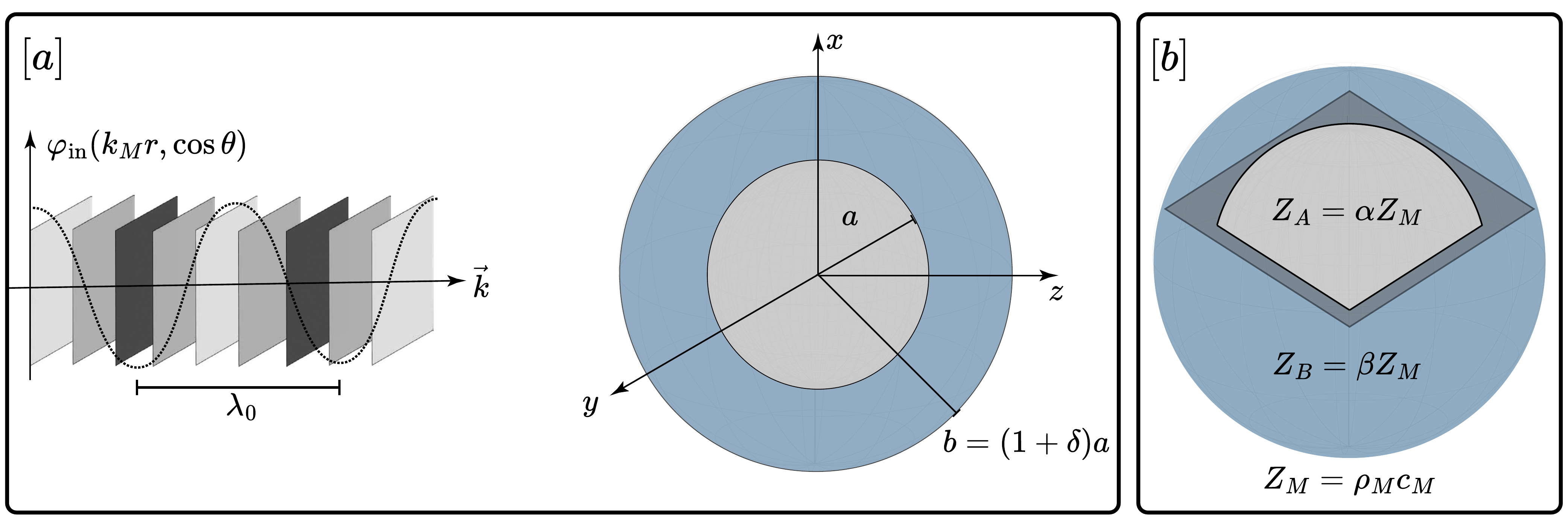}
  \caption{[a] Schematic representation of a monochromatic incident acoustic plane wave $\varphi_{\text{in}}$, with wavelength $\lambda_0$ and wavevector $\vec{k}$, propagating toward a spherical core--shell scatterer, centered in a Cartesian coordinate system $(x,y,z)$. In addition, the core has radius $a$, while the outer radius is $b=(1+\delta)a$. [b] Layered representation of the acoustic impedance distribution. The host medium is characterized by the impedance $Z_M=\rho_M c_M$, whereas the core and shell have impedances written as $Z_A=\alpha Z_M=\rho_A c_A$ and $Z_B=\beta Z_M=\rho_B c_B$, respectively.}
  \label{fig:schematiccoreshell}
\end{figure}
A spherical core--shell system composed of ideal fluids is considered, as illustrated in Fig.~\ref{fig:schematiccoreshell}. The scatterer consists of a core A of radius $a$, surrounded by a shell B with outer radius $b=a(1+\delta)$, where $\delta>0$ is the relative shell thickness. The layered sphere is immersed in a surrounding medium M. Each region $i\in{A,B,M}$ is characterized by its density $\rho_i$, sound speed $c_i$, and acoustic impedance $Z_i=\rho_i c_i$. The impedance contrasts of the core and shell relative to the host are then defined as $Z_A=\alpha Z_M$ and $Z_B=\beta Z_M$. 
Although $\alpha$ and $\beta$ specify the impedance contrasts, they do not determine how each contrast is divided between density and sound speed. Consequently, different combinations of $\rho_i$ and $c_i$ may produce the same values of $\alpha$ and $\beta$. This introduces an additional degree of freedom, allowing physically distinct configurations to share the same impedance contrast. With the material configuration established, the scattering problem is described by a monochromatic plane wave of angular frequency $\omega$ and wavenumber $k_M$ propagating through the host medium toward the structure. In the linear regime, the acoustic field in ideal, irrotational fluids is described by small-amplitude, adiabatic perturbations
~\cite{anderson1950, farra1951}. The dynamics are governed by a scalar potential $\varphi$ whose spatial part satisfies the Helmholtz equation $(\nabla^2+k^2)\varphi(\mathbf{r})=0$, with the wavenumber $k=\omega/c$ for  each respective medium~\cite{morse1948, landau1987}. This potential also determines the physical acoustic fields, with the pressure and velocity field given by $p(\mathbf{r},t)=-\rho\,\partial_t\left[\varphi(\mathbf{r})e^{-i\omega t}\right]$ and $\mathbf{v}(\mathbf{r},t)=\nabla\varphi(\mathbf{r})e^{-i\omega t}$, respectively. Since the scalar potential satisfies the Helmholtz equation in each region and the geometry is spherical, the spatial dependence of the fields is naturally expanded in terms of spherical harmonics~\cite{elliott2024}. In particular, for a plane wave incident along the $z$-axis, the expansion is written as
\begin{subequations}
\begin{align}
\varphi_{\mathrm{in}}(k_M r,\cos\theta)
&= A \sum_{\ell=0}^{\infty} (2\ell+1)\mathrm{i}^\ell j_\ell(k_M r) P_\ell(\cos\theta), \label{eq:phiin} \\
\varphi_B(k_B r,\cos\theta)
&= A \sum_{\ell=0}^{\infty} (2\ell+1)\mathrm{i}^\ell
\left[c_\ell j_\ell(k_B r) + d_\ell y_\ell(k_B r)\right] P_\ell(\cos\theta), \label{eq:phi2} \\
\varphi_A(k_A r,\cos\theta)
&= A \sum_{\ell=0}^{\infty} (2\ell+1)\mathrm{i}^\ell
b_\ell j_\ell(k_A r) P_\ell(\cos\theta), \label{eq:phi1} \\
\varphi_{\mathrm{sc}}(k_M r,\cos\theta)
&= A \sum_{\ell=0}^{\infty} (2\ell+1)\mathrm{i}^\ell
s_\ell h_\ell^{(1)}(k_M r) P_\ell(\cos\theta). \label{eq:phisc}
\end{align}
\end{subequations}
Here, $A$ denotes the amplitude of the incident wave, $j_\ell$ and $y_\ell$ are the spherical Bessel and Neumann functions, respectively, $P_\ell$ are Legendre polynomials, and $h_\ell^{(1)}$ represents the spherical Hankel function of the first kind. The field representations in each region follow from regularity at the origin and the Sommerfeld radiation condition. At fluid--fluid interfaces, the acoustic boundary conditions require the continuity of both acoustic pressure and normal particle velocity. In terms of the scalar velocity potential $\varphi$, these conditions are expressed as $\rho_i \varphi_i = \rho_e \varphi_e$ and $ \partial_r \varphi_i = \partial_r \varphi_e $, where the subscripts $i$ and $e$ denote the media on the inner and outer sides of the interface, respectively. Applying these conditions at the boundaries $r=a$ and $r=b$ yields, for each multipole order $\ell$, a $4 \times 4$ linear system for the unknown coefficients $\mathbf{x}_\ell = [b_\ell, c_\ell, d_\ell, s_\ell]^T$,
\begin{equation}
\label{eq:sistema_direto}
\begin{bmatrix}
0 & -j_\ell(w_B) & -y_\ell(w_B) & \rho_{\text{MB}} h_\ell^{(1)}(w_M) \\
0 & -j_\ell'(w_B) & -y_\ell'(w_B) & k_{\text{MB}} h'^{(1)}_\ell(w_M) \\
-\rho_{\text{AB}} j_\ell(x_A) & j_\ell(x_B) & y_\ell(x_B) & 0 \\
-k_{\text{AB}} j_\ell'(x_A) & j_\ell'(x_B) & y_\ell'(x_B) & 0
\end{bmatrix}
\begin{bmatrix}
b_\ell \\
c_\ell \\
d_\ell \\
s_\ell
\end{bmatrix}
=
\begin{bmatrix}
-\rho_{\text{MB}} j_\ell(w_M) \\
-k_{\text{MB}} j_\ell'(w_M) \\
0 \\
0
\end{bmatrix},
\end{equation}
where $x_A = k_A a$, $x_B = k_B a$, $w_M = k_M b = (1+\delta)x_M$, and $w_B = k_B b$ are the dimensionless arguments, $\rho_{ij} = \rho_i/\rho_j$ represents the density ratios, and $k_{ij} = k_i/k_j = c_j/c_i$ denotes the wave number (or sound speed) ratios.

The analytical expressions for the coefficients can be significantly simplified by invoking the standard Wronskian identities for spherical Bessel functions, i.e., $j_\ell(s)y_\ell'(s) - j_\ell'(s)y_\ell(s) = 1/s^2$ and $h_\ell^{(1)}(s)j_\ell'(s) - h'^{(1)}_\ell(s)j_\ell(s) = -i/s^2$. Evaluating these Wronskians at the arguments $x_B$ and $w_M$ reduces the numerators to purely algebraic terms. By absorbing the boundary Bessel functions directly into the coupling tensor, we define a generalized boundary matrix element $\Gamma_{\mu\nu}^{fg}$ for any adjacent media $\mu, \nu$ and wavefunctions $f,g \in \{j, y, h\}$ as
\begin{equation}
\Gamma_{\mu\nu}^{fg} = \frac{\rho_\mu}{\rho_\nu} f_\ell(s_\mu) g_\ell'(s_\nu) - \frac{c_\nu}{c_\mu} f_\ell'(s_\mu) g_\ell(s_\nu),
\end{equation}
where $s_\mu, s_\nu$ are the arguments at the corresponding interface. The entire system of coefficients reduces to 
\begin{subequations}
\label{eq:solucoes_wronskiano_c}
\begin{align}
b_\ell &= -\frac{i}{w_M^2} \frac{\rho_M}{\rho_B} \frac{c_B}{c_M}\frac{1}{\mathcal{D}_\ell x_B^2}, \\[6pt]
c_\ell &= -\frac{i}{w_M^2} \frac{\rho_M}{\rho_B} \frac{c_B}{c_M}\frac{1}{\mathcal{D}_\ell} \Gamma_{AB}^{jy}, \\[6pt]
d_\ell &= \frac{i}{w_M^2} \frac{\rho_M}{\rho_B} \frac{c_B}{c_M}\frac{1}{\mathcal{D}_\ell} \Gamma_{AB}^{jj}, \\[6pt]
s_\ell &= \frac{1}{\mathcal{D}_\ell} \left( \Gamma_{AB}^{jj}\Gamma_{MB}^{jy} - \Gamma_{AB}^{jy}\Gamma_{MB}^{jj} \right), 
\\
\mathcal{D}_\ell &= \Gamma_{AB}^{jy}\Gamma_{MB}^{hj} - \Gamma_{AB}^{jj}\Gamma_{MB}^{hy}.
\end{align}
\end{subequations}
The denominator $\mathcal{D}_\ell$ captures the dynamic interaction between the core and the shell. Mediated by the impedance contrasts, the shell may act as a spherical Fabry-Pérot resonator, coupling the quasi-normal modes of both regions. In this coupled regime, rather than localizing in a single medium, the acoustic field continuously exchanges energy between the core and the shell prior to outward radiation.

The formulation must recover the homogeneous-sphere when the distinction between core and shell disappears. If the shell becomes indistinguishable from the core ($B\to A$, $\rho_A/\rho_B\to1$, $c_B/c_A\to1$, $x_B\to x_A$), the inner-interface factor vanishes identically, $j_\ell(x_A)j_\ell'(x_A)-j_\ell'(x_A)j_\ell(x_A)=0$, while the second factor reduces to the Wronskian $j_\ell(x_A)y_\ell'(x_A)-j_\ell'(x_A)y_\ell(x_A)=1/x_A^{2}$.  $\mathcal{D}_\ell$ then reproduces the denominator of a homogeneous sphere of radius $b$ and material $A$ in $M$. Conversely, if the shell becomes indistinguishable from the host ($B\to M$, $\rho_M/\rho_B\to1$, $c_B/c_M\to1$, $w_B\to w_M$), the outer-interface factors combine via the same Wronskian. Using $h_\ell^{(1)}=j_\ell+i y_\ell$, the shell contributions merge into a single outgoing-wave term, giving the denominator for a homogeneous sphere of radius $a$ (core $A$) embedded directly in $M$, giving
\begin{equation}
\mathcal{D}_\ell
\propto
j_\ell(x_A)h'^{(1)}_\ell(x_M)
-
\dfrac{Z_M}{Z_A}j_\ell'(x_A)h_\ell^{(1)}(x_M),
\end{equation}
up to an overall nonzero factor. This is precisely the single-interface denominator for a homogeneous sphere made of material $A$ in medium $M$. This limiting behavior is consistent with previous results for concentric fluid spheres, where the solution was shown to recover Anderson's single-fluid-sphere result when the two radii coincide and the layered structure reduces to a homogeneous sphere~\cite{mcnew2009}.

From the resulting fields, the total acoustic energy is defined as
\begin{equation}
W_T
=
\frac{\rho}{4}
\int_V
\left|\nabla\varphi\right|^2\,dV
+
\frac{\rho k^2}{4}
\int_V
\left|\varphi\right|^2\,dV .
\end{equation}
For the potential energy, $W_P^{\text{core}}$, the spherical-harmonic expansion of $\varphi_A$ and the orthogonality of the Legendre polynomials give
\begin{equation}
\frac{2f_0 W_P^{\text{core}}}{Z_A|A|^2}
=
\frac{x_A^3}{2}
\sum_{\ell=0}^{\infty}
|b_\ell|^2(2\ell+1)
\left[
j_\ell^2(x_A)
-
j_{\ell-1}(x_A)j_{\ell+1}(x_A)
\right],
\end{equation}
with $Z_A = \rho_A c_A$, and $b_\ell$ the internal expansion coefficients~\cite{rodrigues2026}, it remains to evaluate the kinetic energy. Instead of expanding $|\nabla\varphi_A|^2$ into radial and angular derivative terms, as done in Ref.~\cite{rodrigues2026}, this term can be rewritten through the identity $\nabla\cdot(\varphi_A^*\nabla\varphi_A) = |\nabla\varphi_A|^2 + \varphi_A^*\nabla^2\varphi_A$. Since $\nabla^2\varphi_A = -k_A^2\varphi_A$, we obtain $|\nabla\varphi_A|^2 = \nabla\cdot(\varphi_A^*\nabla\varphi_A) + k_A^2|\varphi_A|^2$. The divergence theorem then turns the volume integral of the divergence into a boundary term evaluated at $r=a$.
\begin{equation}
\frac{2f_0 W_K^{\text{core}}}{Z_A|A|^2}
=
\sum_{\ell=0}^{\infty}
|b_\ell|^2(2\ell+1)
\left\{
\frac{x_A^3}{2}
\left[
j_\ell^2(x_A)
-
j_{\ell-1}(x_A)j_{\ell+1}(x_A)
\right]
+
x_A^2j_\ell(x_A)j_\ell'(x_A)
\right\}.
\label{eq:newcore}
\end{equation}
This route avoids the radial and angular derivative expansion of the velocity field and removes the numerical integrations required in the standard formulation.

The same procedure can be applied to the shell ($a \leq r \leq b$). For the potential contribution as done in Ref.~\cite{alcaras2023}, the change of variable $x = k_B r$ is introduced, so that the radial limits become $x_B$ and $w_B$. The angular part is then evaluated using the orthogonality relation of the Legendre polynomials, reducing the double modal sum to a single sum over $\ell$. The remaining radial dependence is given by products of spherical Bessel and Neumann functions, which are real for real arguments; hence the cross terms simplify to $\mathrm{Re}(c_\ell^* d_\ell)$.

The required radial integrals are written in terms of the boundary functions
\begin{equation}
\begin{split}
F_{jj}(x_B,w_B)
&=
\left.\left[
x^3
\left(
j_\ell^2(x)
-
j_{\ell-1}(x)j_{\ell+1}(x)
\right)
\right]\right|_{x_B}^{w_B},
\\
F_{yy}(x_B,w_B)
&=
\left.\left[
x^3
\left(
y_\ell^2(x)
-
y_{\ell-1}(x)y_{\ell+1}(x)
\right)
\right]\right|_{x_B}^{w_B},
\\
F_{jy}(x_B,w_B)
&=
\left.\left[
\frac{x^3}{2}
\left(
2j_\ell(x)y_\ell(x)
-
j_{\ell+1}(x)y_{\ell-1}(x)
-
j_{\ell-1}(x)y_{\ell+1}(x)
\right)
\right]\right|_{x_B}^{w_B}.
\end{split}
\end{equation}
Therefore, the potential energy stored in the shell is obtained as
\begin{equation}
\frac{2f_0W_P^{^{\text{shell}}}}{Z_B|A|^2}
=
\sum_{\ell=0}^{\infty}
(2\ell+1)
\left[
\frac{|c_\ell|^2}{2}F_{jj}(x_B,w_B)
+
\frac{|d_\ell|^2}{2}F_{yy}(x_B,w_B)
+
\mathrm{Re}\left(c_\ell^*d_\ell\right)F_{jy}(x_B,w_B)
\right].
\label{eq:WP_B_final}
\end{equation}

Evaluating the kinetic contribution proceeds analogously through the gradient of $\varphi_B$, which introduces both radial and angular components. By applying integration by parts alongside the spherical Bessel differential equation, the radial derivative integrals are systematically recast in terms of boundary values and the previously defined functions $F_{jj}$, $F_{yy}$, and $F_{jy}$. Consequently, the final expression for the kinetic energy is obtained straightforwardly
\begin{equation}
\begin{split}
\frac{2f_0W_K^{\text{shell}}}{Z_B|A|^2}
&=
\frac{2f_0W_P^{\text{shell}}}{Z_B|A|^2}
+
\sum_{\ell=0}^{\infty}
(2\ell+1)
\Bigg\{
|c_\ell|^2
\left.\left[
x^2j_\ell(x)j_\ell'(x)
\right]\right|_{x_B}^{w_B}
\\
&\qquad
+
|d_\ell|^2
\left.\left[
x^2y_\ell(x)y_\ell'(x)
\right]\right|_{x_B}^{w_B}
+
2\mathrm{Re}\left(c_\ell^*d_\ell\right)
\left.\left[
x^2j_\ell(x)y_\ell'(x)
\right]\right|_{x_B}^{w_B}
\Bigg\}.
\end{split}
\label{eq:WK_B_final}
\end{equation}
In practice, the mixed terms proportional to $\mathrm{Re}(c_\ell^*d_\ell)$ capture the spatial modulation of the field within the finite shell. The superposition of regular ($j_\ell$) and singular ($y_\ell$) radial functions provides the exact degrees of freedom required to simultaneously satisfy the continuity boundary conditions at both interfaces ($r=a$ and $r=b$), governing the spatial distribution of the confined acoustic energy.

\section{results}
\label{sec:results}

\begin{figure}[htb]
    \centering
    \includegraphics[width=0.9\columnwidth]{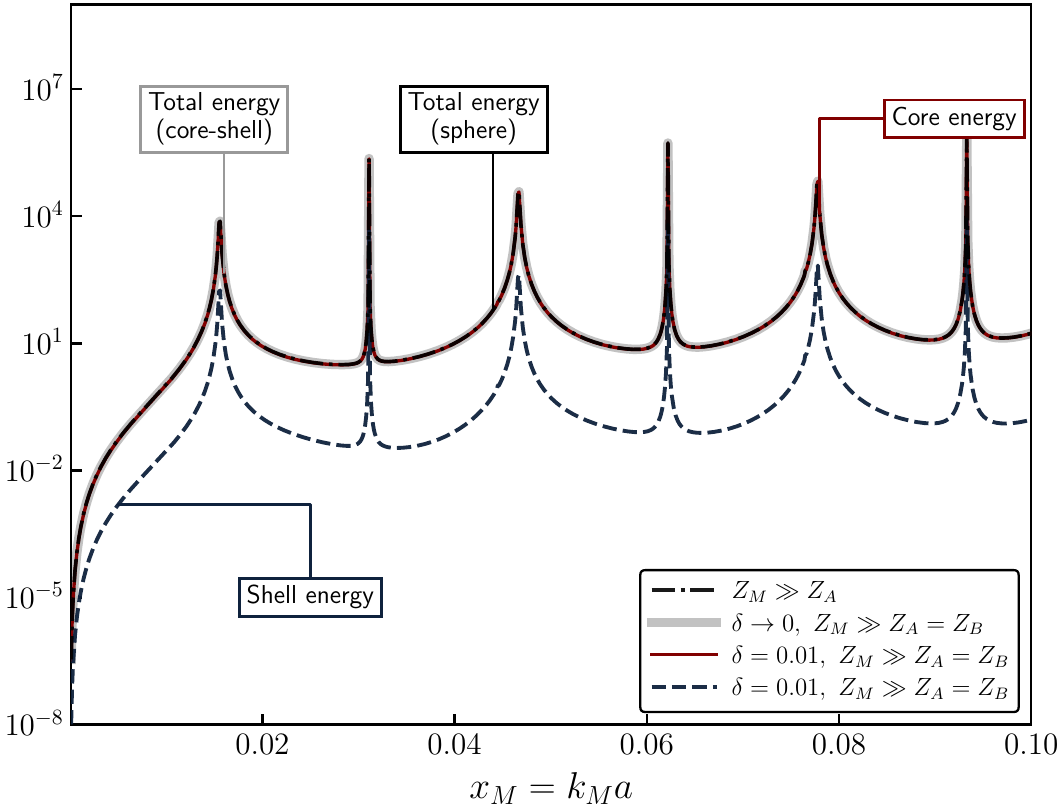}
    \caption{\label{fig:wtorder}Normalized internal energy as a function of the dimensionless size parameter $x_M=k_Ma$ for $\alpha=\beta=0.01$. The core--shell formulation in the vanishing-shell limit, $\delta\rightarrow0$, is compared with the independent numerical solution for a single fluid sphere. The remaining curves show the internal-energy response for finite shell thicknesses $\delta$.}
\end{figure}
As a preliminary comparison, Fig.~\ref{fig:wtorder} compares a core-shell particle (with inner radius $a$ and outer radius $b=(1+\delta)a$, for $\delta=0.01$) to a homogeneous sphere of radius $a$. Notably, the analytical response of the sphere, derived from Eq.~\eqref{eq:newcore}, reproduces the results of Ref.~\cite{rodrigues2026} without requiring numerical integration. To examine the combined material and geometric effects, water is taken as the surrounding medium, with $\rho_M=1024~\mathrm{kg/m^{3}}$ and $c_M=1522~\mathrm{m/s}$, whereby its impedance $Z_M$ is used as the reference. Subsequently, the internal impedances are written as $Z_A$ and $Z_B=$. Since $Z_i=\rho_i c_i$, it follows that the same impedance contrast can arise through entirely different physical mechanisms. Specifically, variations in $c_i$, apart from changing the velocity, inherently alter the internal wavelengths; conversely, variations in $\rho_i$ modify only the boundary conditions at the interfaces. While homogeneous spheres typically require a large impedance mismatch ($Z_M\gg Z_A$) to resonate, the additional interface in the core-shell structure introduces an interplay among $Z_A$, $Z_B$, and $Z_M$, which may allow further resonant configurations.
\begin{figure*}[htb]
    \centering
    
    \begin{minipage}[t]{0.325\textwidth}
        \centering
        \includegraphics[width=\linewidth]{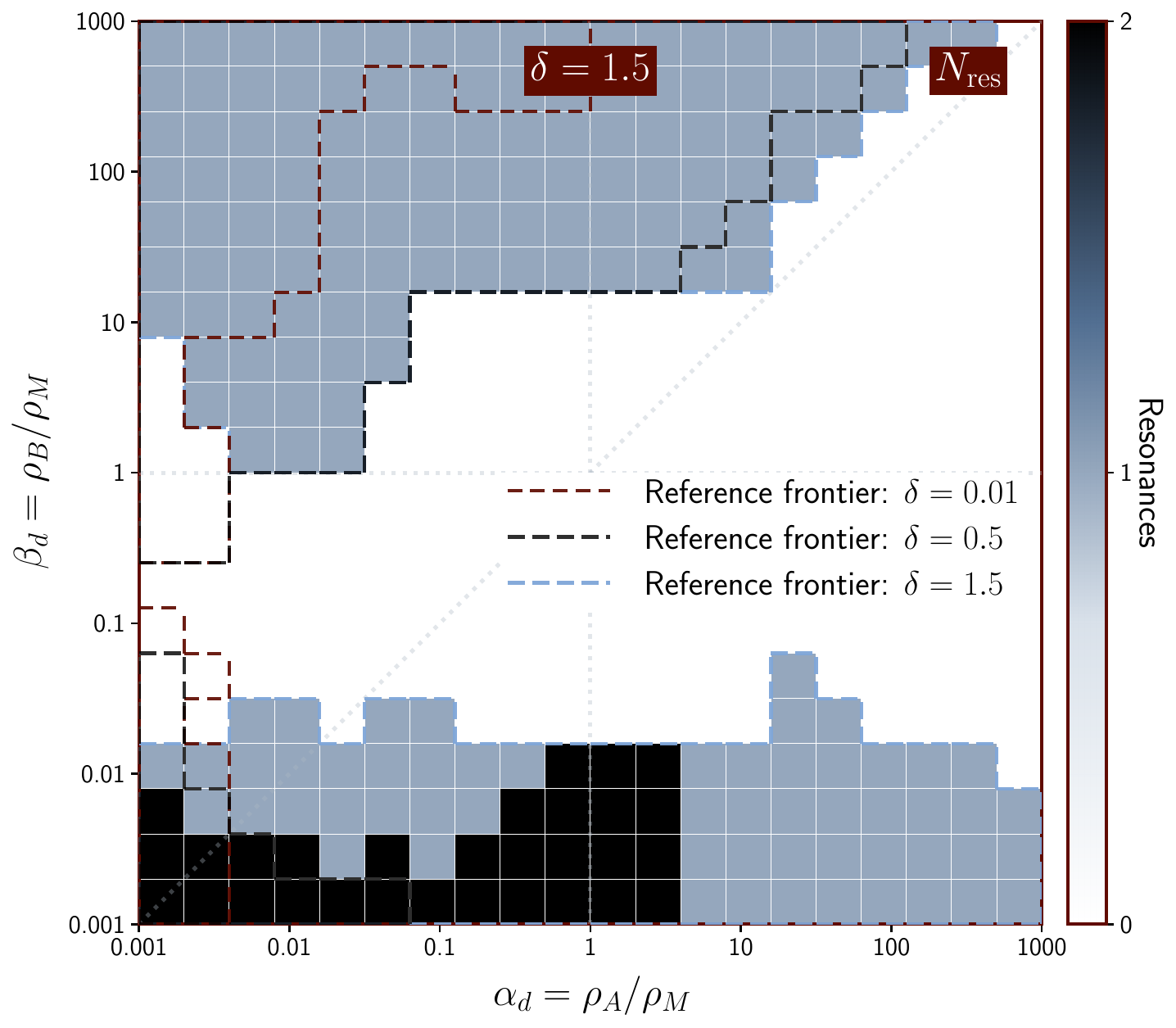}
    \end{minipage}
    \hfill
    \begin{minipage}[t]{0.325\textwidth}
        \centering
        \includegraphics[width=\linewidth]{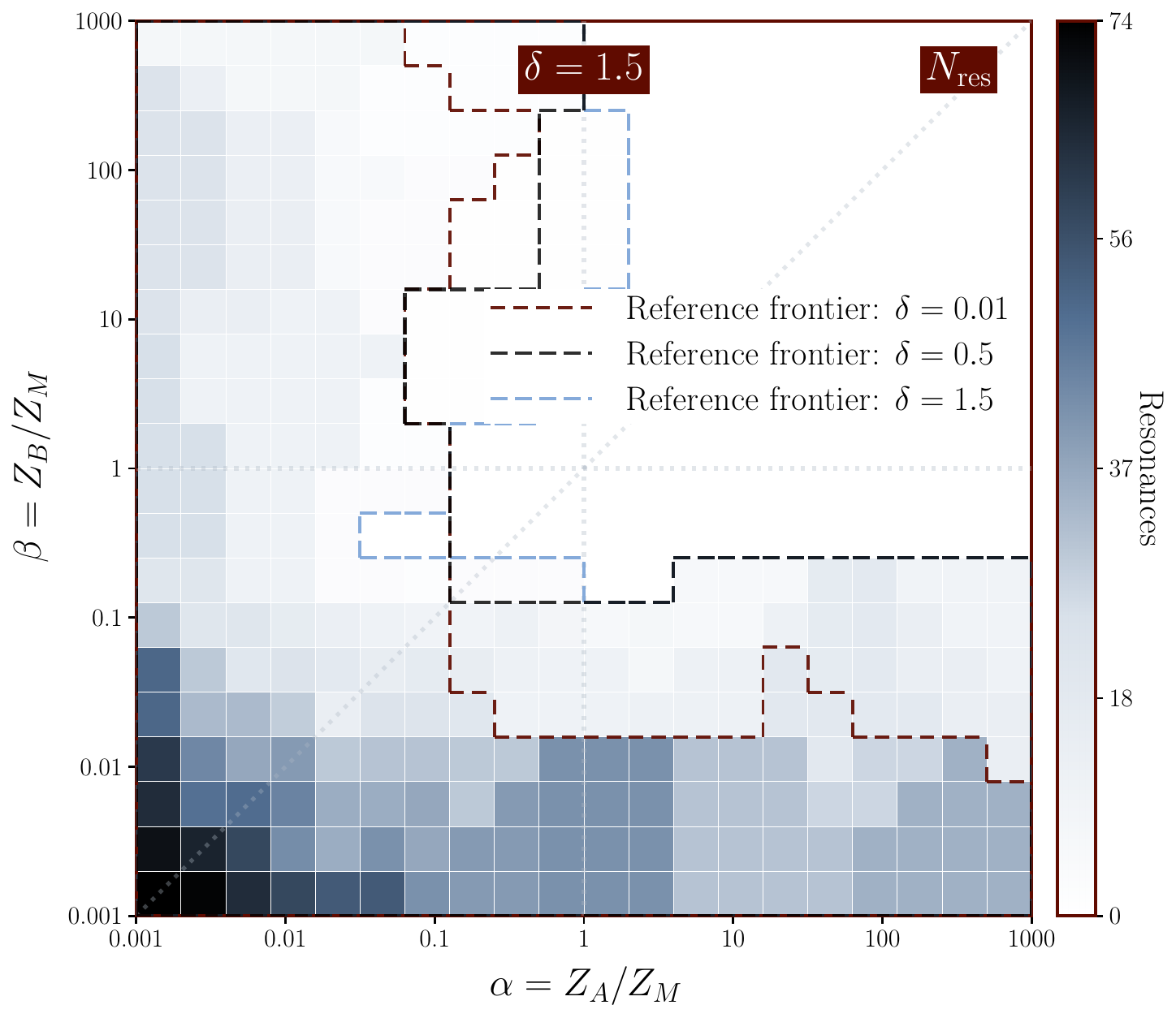}
    \end{minipage}
    \hfill
    \begin{minipage}[t]{0.325\textwidth}
        \centering
        \includegraphics[width=\linewidth]{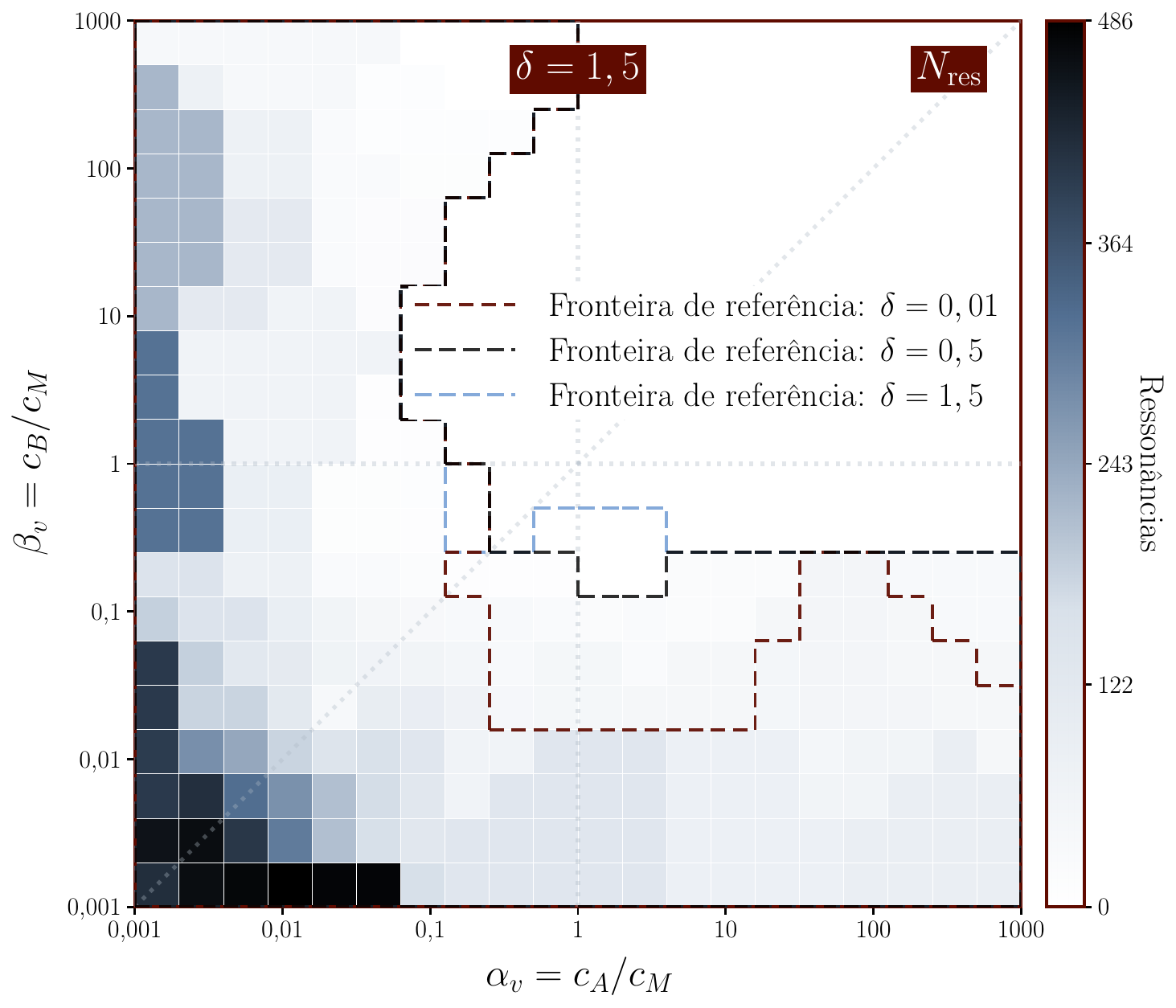}
    \end{minipage}

\caption{\label{fig:max_resonances_contrast}
Resonance maps in the impedance plane $(\alpha,\beta)$ showing the maximum number of acoustic resonances per bin up to $x_M=1$ for density-driven, mixed, and sound-speed-driven contrasts, from left to right. All panels use the same scales for direct comparison. The colormap corresponds to $\delta=1.5$, while the overlaid contours show the shift of the non-resonant boundary for other shell thicknesses. The mixed case uses $f_{\mathrm{mix}}=0.5$.}
\end{figure*}

The $(\alpha,\beta)$ plane in Fig.~\ref{fig:max_resonances_contrast} represents the acoustic impedances of the core and shell relative to the host. Since the same impedance contrast can be obtained through different combinations of density and sound speed, three realizations are considered: density-driven contrasts with fixed sound speeds, sound-speed-driven contrasts with fixed densities, and mixed contrasts in which density and sound speed contribute equally. For each case, the colored contours delimit the non-resonant domain for different shell-thickness ratios. The results show that introducing a finite shell extends the resonant region into impedance configurations that are non-resonant in the nearly vanishing-shell limit.

However, the effect of increasing $\delta$ depends on the origin of the contrast. For density-driven and mixed impedance ratios, the non-resonant domain contracts appreciably as the shell becomes thicker, producing a relevant enlargement of the resonant regions. In the mixed case, with $f_{\mathrm{mix}}=0.5$, the core and shell impedance contrasts are equally distributed between density and sound speed. By contrast, for sound-speed-driven case, the presence of the shell is essential for expanding the resonant state space, but this expansion is already largely established for $\delta=0.1$. Consequently, the boundary obtained for $\delta=1.5$ remains nearly unchanged relative to that for $\delta=0.1$, indicating that further increases in shell thickness have a minor impact on the ressonance frequency.

Moreover, the impedance hierarchy governs the mixed and sound-speed maps in a closely similar manner. If the host has the lowest impedance $(Z_M<Z_B<Z_A)$ or $(Z_M<Z_A<Z_B)$, the system stays in the non-resonant regime. But if either the core or the shell drops below $Z_M$, resonances start to appear. This behavior reveals a notable symmetry: the mixed and sound-speed regimes are invariant under the exchange of core and shell ($\alpha \leftrightarrow \beta$), because the impedance contrasts in these cases are determined solely by the sound-speed ratios, which can be interchanged without altering the internal energy distribution for a fixed outer radius. As a consequence, a forbidden quadrant emerges when the impedances of both core and shell exceed that of the host. In this region, the system remains strictly non-resonant, regardless of the shell thickness. The existence of such a forbidden quadrant was first identified for a homogeneous fluid sphere, where the condition $Z_{\mathrm{sphere}} > Z_{\mathrm{host}}$ suppresses internal resonances~\cite{simao2016tunneling}. The present work generalizes this concept to the layered core--shell geometry, showing that the same impedance hierarchy principle operates, now governed by the pair $(\alpha,\beta)$.

The density-driven case works differently. The resonant range grows with $Z_B/Z_M$. The density map does not possess the $\alpha \leftrightarrow \beta$ symmetry: the geometric asymmetry introduced by the shell thickness breaks the invariance under material exchange, meaning that the condition for resonance is sensitive to which material forms the shell and which forms the core. Hence, no simple forbidden quadrant exists for density-driven impedance contrast.

Once the role of the impedance hierarchy in governing the resonant domains has been established, the specific behavior associated with the sound-speed-driven, mixed, and density-driven regimes can be examined separately. A detailed discussion of these regimes, including the influence of shell thickness on their spectral structure is provided in the Appendix~\ref{sec:velocity}--~\ref{sec:density}. The systematic organization revealed by the resonance maps is then used to derive compact empirical relations for resonance existence and recurrence.

\subsection{Empirical relations for resonance design}

The resonance maps were constructed by scanning, for each contrast regime, a broad grid of impedance ratios together with different shell thicknesses. For each $(\alpha,\beta,\delta)$ configuration, the internal energy spectrum was evaluated to determine whether a resonance occurs and, when present, its spectral position. The systematic dependence observed throughout the parameter space allows the graphical information contained in the maps to be expressed through explicit empirical relations. Two complementary aspects of the resonant response are considered: the conditions governing if there is a resonance or not and the recurrence relation describing the sequence of resonant positions.

To this end, two complementary empirical relations were built directly from the numerically resolved resonance dataset spanning the three regimes. The first is a logistic regression that predicts whether a resonance exists at all, given  $\alpha$, $\beta$, and $\delta$, the probability $P_{\mathrm{res}}(\alpha,\beta,\delta) =(1+\textrm{e}^{-g(\alpha,\beta,\delta)})^{-1}$, where
\begin{equation} 
g(\alpha,\beta,\delta) = (c_0 + c_3\delta + c_9\delta^2) + \ln\alpha  (c_1 + c_6\delta + c_4\ln\alpha + c_5\ln\beta) + \ln\beta  (c_2 + c_8\delta + c_7\ln\beta) 
\label{eq:g-function}
\end{equation}
with resonance predicted for $P_{\mathrm{res}} \geq 0.5$, equivalently $g \geq 0$.
\begin{table}[ht]
\centering
\caption{Coefficients of the logistic regression for resonance, following Eq.~\eqref{eq:g-function}, across the three parametric regimes.}
\label{tab:coefs-master}
\begin{ruledtabular}
\begin{tabular}{lccc}
\textbf{Coeff} & \textbf{Density Contrast} & \textbf{Velocity Contrast} & \textbf{Mixed Contrast} \\
\midrule
$c_{0}$ & $-8.\mathbf{86} \pm 0.31$ & $-4.\mathbf{77} \pm 0.27$ & $-7.\mathbf{17} \pm 0.34$ \\
$c_{1}$ & $-1.6\mathbf{87} \pm 0.076$ & $-1.6\mathbf{24} \pm 0.091$ & $-1.3\mathbf{78} \pm 0.078$ \\
$c_{2}$ & $+0.8\mathbf{28} \pm 0.053$ & $-1.2\mathbf{15} \pm 0.089$ & $-1.0\mathbf{84} \pm 0.082$ \\
$c_{3}$ & $+3.\mathbf{74} \pm 0.50$ & $+2.\mathbf{55} \pm 0.65$ & $+4.\mathbf{81} \pm 0.64$ \\
$c_{4}$ & $+0.00\mathbf{04} \pm 0.0073$ & $+0.1\mathbf{88} \pm 0.022$ & $+0.1\mathbf{42} \pm 0.015$ \\
$c_{5}$ & $-0.2\mathbf{24} \pm 0.012$ & $-0.1\mathbf{92} \pm 0.024$ & $-0.2\mathbf{35} \pm 0.020$ \\
$c_{6}$ & $+0.6\mathbf{21} \pm 0.064$ & $+0.\mathbf{13} \pm 0.13$ & $+0.5\mathbf{71} \pm 0.100$ \\
$c_{7}$ & $+0.3\mathbf{20} \pm 0.012$ & $+0.3\mathbf{35} \pm 0.020$ & $+0.3\mathbf{35} \pm 0.017$ \\
$c_{8}$ & $-0.2\mathbf{83} \pm 0.056$ & $-0.\mathbf{67} \pm 0.11$ & $-0.4\mathbf{26} \pm 0.093$ \\
$c_{9}$ & $+0.\mathbf{00} \pm 0.29$ & $-1.\mathbf{41} \pm 0.44$ & $-2.\mathbf{33} \pm 0.41$ \\
\end{tabular}
\end{ruledtabular}
\end{table}

The coefficients, listed in Table~\ref{tab:coefs-master}, were fitted independently for each contrast regime. The classifier performs strongly in all three cases, though with some variation that mirrors the resonance statistics discussed earlier: for the sound-speed-driven regime, it reaches a test accuracy of $99.04\%$ (precision $99.64\%$, recall $98.08\%$, ROC-AUC $0.9997$); for the mixed regime, a test accuracy of $97.57\%$ (precision $96.95\%$, recall $96.36\%$, ROC-AUC $0.9982$); and for the density-driven regime, a test accuracy of $95.37\%$ (precision $90.96\%$, recall $78.80\%$, ROC-AUC $0.9837$). The comparatively lower recall in the density-driven case reflects the scarcity of resonant configurations in this regime (positive rate of only $\sim16\%$, against $\sim36$--$42\%$ for the other two), which makes the minority resonant class intrinsically harder to identify, yet the model still recovers the large majority of resonant cases and maintains a high overall discrimination power. Not all coefficients carry equal weight, however: $c_4$ and $c_9$ are not statistically significant in the density-driven regime, and $c_6$ is not significant in the sound-speed-driven regime, though the full ten-term expression is retained uniformly for comparability.

The second relation addresses the resonant configurations themselves: once a resonance is known to exist, successive resonance positions follow a smooth recurrence as functions of $\eta =(Z_A-Z_B)/(Z_A+Z_B) $ and $\delta$,
\begin{subequations}
\begin{align}
\label{eq:recorrencia}
x_{M,n+1} &= x_{\mathrm{ref}}A(\eta,\delta)+B(\eta,\delta)x_{M,n}, \\
A(\eta,\delta) &= a_0 + a_1\,\eta + a_2\,\delta + a_3\,\eta^2 + a_4\,\eta\delta + a_5\,\delta^2, \\
B(\eta,\delta)  &= b_0 + b_1\,\eta + b_2\,\delta + b_3\,\eta^2 + b_4\,\eta\delta + b_5\,\delta^2.
\end{align}
\end{subequations}
This recurrence relation was obtained by fitting the second-order polynomial functions in Eq.~\eqref{eq:recorrencia} to the successive resonance positions, with coefficients reported in Table~\ref{tab:coefs-recurrence}. In the sound-speed-driven regime, built from 2826 resonant curves out of 6800 anchored configurations, the fit achieves $R^2 = 0.9989$ on the test set (RMSE $= 0.0094$, MAE $= 0.0034$, MAPE $= 1.32\%$). In the mixed regime, built from 2486 resonant curves out of the same 6800 anchored configurations, the fit is somewhat less tight but still strong, with $R^2 = 0.9812$ on the test set (RMSE $= 0.0356$, MAE $= 0.0238$, MAPE $= 6.51\%$), reflecting the sparser and less regular resonance pattern already noted for this regime. The density-driven regime was not included in this recurrence relation, since its far smaller and less regular set of resonant configurations does not support a stable low-order polynomial fit; in this regime, resonance is instead well captured by the logistic classifier alone. In both fitted regimes, the anchoring configuration was defined from the homogeneous-sphere limit, using $x_{\mathrm{ref}} = \sqrt{{3}/{m\,m_t}}$, $m = c_M/c_A$, $m_t = Z_M/Z_A=1/\alpha$,
ensuring that the fitted recurrence remains consistent with the known single-material resonance condition as $\delta$ varies.

\begin{table}[ht]
\centering
\caption{Recurrence relation coefficients for $A(\eta,\delta)$ and $B(\eta,\delta)$, following Eq.~\eqref{eq:recorrencia}, in the velocity-driven and mixed regimes.}
\label{tab:coefs-recurrence}
\begin{ruledtabular}
\begin{tabular}{l cc cc}
& \multicolumn{2}{c}{\textbf{Velocity Contrast}} & \multicolumn{2}{c}{\textbf{Mixed Contrast}} \\
\cmidrule(lr){2-3} \cmidrule(lr){4-5}
\textbf{Coeff.} & $A$ & $B$ & $A$ & $B$ \\
\midrule
$0$ &
$+0.44\mathbf{45} \pm 0.0012$ &
$+1.006\mathbf{60} \pm 0.00010$ &
$+1.10\mathbf{38} \pm 0.0087$ &
$+1.02\mathbf{59} \pm 0.0011$ \\

$1$ &
$-0.46\mathbf{28} \pm 0.0013$ &
$+0.010\mathbf{29} \pm 0.00013$ &
$-1.14\mathbf{42} \pm 0.0085$ &
$+0.05\mathbf{74} \pm 0.0013$ \\

$2$ &
$-0.17\mathbf{52} \pm 0.0012$ &
$-0.009\mathbf{14} \pm 0.00029$ &
$-0.41\mathbf{66} \pm 0.0087$ &
$-0.06\mathbf{19} \pm 0.0028$ \\

$3$ &
$+0.018\mathbf{39} \pm 0.00084$ &
$+0.003\mathbf{25} \pm 0.00012$ &
$+0.04\mathbf{12} \pm 0.0057$ &
$+0.02\mathbf{23} \pm 0.0011$ \\

$4$ &
$+0.17\mathbf{51} \pm 0.0012$ &
$-0.009\mathbf{30} \pm 0.00014$ &
$+0.41\mathbf{59} \pm 0.0087$ &
$-0.04\mathbf{39} \pm 0.0013$ \\

$5$ &
$+0.00005\mathbf{22} \pm 0.0000021$ &
$+0.004\mathbf{31} \pm 0.00019$ &
$+0.0002\mathbf{03} \pm 0.000063$ &
$+0.02\mathbf{90} \pm 0.0017$ \\
\end{tabular}
\end{ruledtabular}
\end{table}

Taken together, Eqs.~\eqref{eq:g-function}--\eqref{eq:recorrencia} offer a compact, closed-form route from a target operating point $(\eta,\delta)$ to the impedance ratios most likely to sustain a resonance, and from a candidate material pair $(\alpha,\beta,\delta)$ to a direct estimate of whether that configuration will resonate at all. This is precisely the kind of tunability that matters for metamaterial design: rather than scanning the full $(\alpha,\beta,\delta)$ space numerically for every new application, the resonant region to be located analytically, narrowing the search to a small set of promising core-shell material combinations before any full-spectrum calculation is performed.

\begin{figure}[htb]
    \centering
\includegraphics[width=0.9\columnwidth]{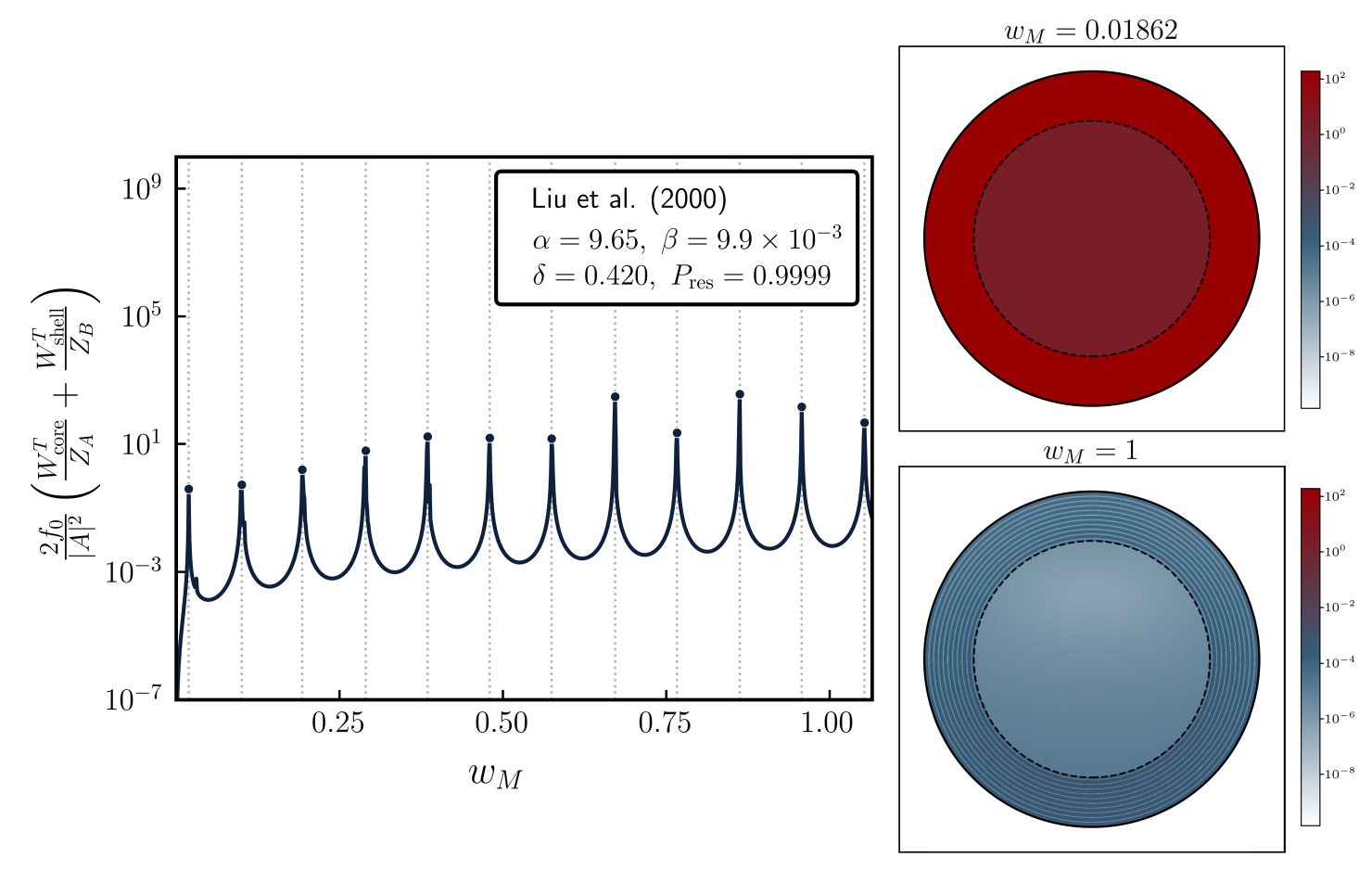}
\caption{\label{fig:liu_field}
Validation against the locally resonant model of Liu \textit{et al.}~\cite{liu2000} (lead core, silicone-rubber shell, epoxy host). The system is modeled using $\rho_A=11.6, \rho_B=1.3, \rho_M=1.18$ (in $10^{3}~\mathrm{kg/m^3}$) and wave speeds $c_A\approx2493, c_B\approx22.9, c_M\approx2540~\mathrm{m/s}$. This strong impedance contrast gives $\alpha\approx9.65$, $\beta\approx0.00992$, and $\delta\approx0.420$ ($\eta>0$ regime). \textit{Left:} Internal acoustic energy spectrum vs.~$w_M=(1+\delta)x_M$; Eq.~$P_{\mathrm{res}}$ predicts this configuration as resonant ($P_{\mathrm{res}}=0.9999$). \textit{Right:} Internal field intensity $|\varphi|^2$ (log scale) at the first resolved resonance ($w_M=0.0186$, top) and at $w_M=1$ (bottom). Dashed/solid circles indicate the core/shell boundaries.}
\end{figure}
\begin{figure}[htb]
    \centering
    \includegraphics[width=0.9\columnwidth]{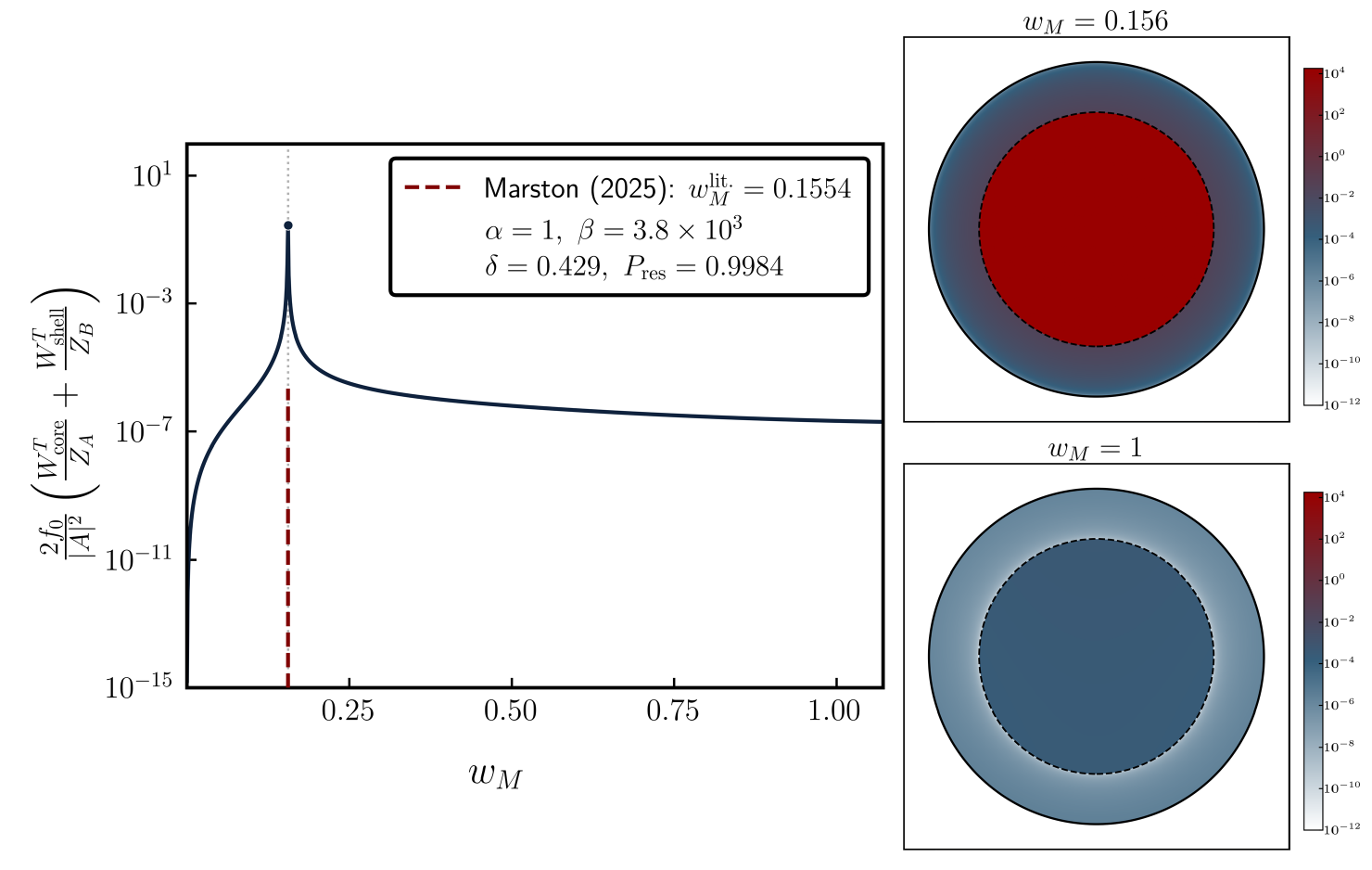}
    \caption{\label{fig:marston_field}Validation against the fluid-sphere system of Marston~\cite{marston2025} (air core, glycerol-solution shell, air host). Since the core and host are identical, $Z_A/Z_M=1$. The shell properties yield $Z_B/Z_M\approx3804$, with a radius ratio $\delta\approx0.429$, placing the system in the $\eta<0$ region. \textit{Left:} Internal acoustic energy spectrum vs.~$w_M=(1+\delta)x_M$. The classifier of $P_{\mathrm{res}}$ predicts resonance ($P_{\mathrm{res}}=0.9984$), with the resolved peak matching the literature value $w_M^{\mathrm{lit.}}=0.1554$ from Ref.~\cite{marston2025}. \textit{Right:} Cross-sectional internal field intensity $|\varphi|^2$ (log scale) near resonance ($w_M=0.156$, top) and off-resonance ($w_M=1$, bottom). Dashed and solid circles mark the core--shell and shell--host boundaries, respectively.}
\end{figure}

Figures~\ref{fig:liu_field} and~\ref{fig:marston_field} illustrate what this validation adds beyond a single number. The Liu system was originally reported as a locally resonant sonic \textit{material} -- a slab embedded with coated lead spheres, whose overall band-gap behavior was measured, but whose microscopic resonant mechanism inside a single scatterer was described only through a simplified, lumped mass-in-mass-spring picture, more illustrative than rigorous. It is also worth stressing that the underlying system is not an ideal fluid at all: the lead core and the silicone-rubber shell are elastic solids, and the effective-fluid treatment adopted here, restricted to the shell's longitudinal wave speed, is itself an approximation to that solid-solid problem. Even so, once cast into the $(\alpha,\beta,\delta)$ language of the present framework, the same impedance mismatch responsible for the mass-spring picture in the original work reproduces the dense, multiplet-rich resonance ladder seen in Fig.~\ref{fig:liu_field}, correctly identified by $P_{\text{res}}$ as strongly resonant, with all peaks marked by the gray line predicted by the Eq.~\eqref{eq:recorrencia}. The internal-field panels make the physical content of this behavior explicit: near the first resolved peak ($w_M=0.0186$), the field fills both core and shell almost uniformly at high amplitude, characteristic of a low-order, quasi-static response, whereas at $w_M=1$ the field organizes into a dense set of concentric standing-wave rings spanning many multipole orders, at markedly lower amplitude -- the spatial signature of the higher-order members of the resonance ladder already seen in the spectrum.

The Marston system, by contrast, is exactly the regime for which the present formalism was built: an ideal-fluid core, shell, and host, with no solid-shell approximation required. Here $P_{\text{res}}$ again predicts a resonance with high confidence, and the energy spectrum in Fig.~\ref{fig:marston_field} reproduces, at essentially the same $w_M$, the single monopole resonance obtained independently in Ref.~\cite{marston2025} from the phase-shift condition $\delta_0=90^{\circ}$. The internal field confirms the physical picture behind that result: near resonance ($w_M=0.156$) the field is sharply concentrated inside the core, consistent with the breathing motion of the air bubble driving the surrounding liquid shell, in direct correspondence with the generalized Minnaert mechanism described in that work; at $w_M=1$, far from resonance, the same core-concentrated pattern persists but at several orders of magnitude lower amplitude, showing that the shell stores comparatively little energy outside the narrow resonant condition -- the opposite spectral behavior from the Liu system, and a direct visual counterpart to the sparse, weakly dependent on $\delta$.

\section{Conclusion}
\label{sec:conclusions}

In this work, we developed a formulation to calculate the acoustic energy stored inside a fluid core--shell sphere using internal acoustic fields. Building on this formulation, the main result of this work is the construction of a {resonance phase diagram}, which maps the resonant and non-resonant regions of the parameter space for different types of material; and two empirical relations that predicts resonant regimes from material properties via $P_{\textrm{res}}$ and the recurrent relation Eq.~\eqref{eq:recorrencia}.

To isolate how internal wave propagation and boundary mismatches independently govern acoustic resonance, three contrast regimes were examined: velocity-driven (fixed densities, altering internal wavelengths), density-driven (fixed sound speeds, altering interface boundary conditions), and mixed-driven (balanced velocity and density contributions). The resonant landscape is compactly defined by the core and shell impedance contrasts relative to the host $\alpha = Z_A/Z_M$ and $\beta = Z_B/Z_M$, the relative shell thickness $\delta$, and the core-shell impedance asymmetry $\eta = (\alpha-\beta)/(\alpha+\beta)$. In velocity- and mixed-driven regimes, the response exhibits core-shell symmetry $\alpha \leftrightarrow \beta$ and a non-resonant quadrant ($\alpha, \beta > 1$), where positive asymmetry ($\eta > 0$) allows shell thickness ($\delta$) to serve as a key frequency-tuning parameter. In contrast, the density-driven case breaks the $\alpha \leftrightarrow \beta$ symmetry and produce resonances even for ($\alpha, \beta > 1$), signaling a substantial shift from the velocity case and that the resonant scattering cannot be summarized by the impendances alone. 

Even though the velocity-, mixed-, and density-drive regimes presents different structures, the global sequence of resonant peaks can still be described by a
single recurrence relation Eq.~\eqref{eq:recorrencia}. This universal behavior allows one to predict the resonant behavior for any given material combination. 
Together, our findings turn the phase diagram from a descriptive map into a predictive design tool: given a target operating point, one can locate the impedance ratios most likely to sustain a resonance, or, given a candidate pair of core and shell materials, decide whether that specific combination will resonate before any full spectral calculation is carried out.

The value of this predictive picture is in excellent agreement with two independent, experimentally realized systems from the literature from the locally resonant sonic material of Liu \textit{et al.}~\cite{liu2000}, and the concentric fluid-sphere system recently analyzed by Marston~\cite{marston2025}. In both cases, our findings correctly identifies the resonant behavior observed in the original works. Despite these advances for metamaterial designs, our formulation neglect losses entirely via viscosity, thermal conduction, and deformations. Fully accounting for energy losses in future designs will require incorporating viscoelastic properties, thermal and mechanical expansion, as well as mode conversion between longitudinal and transverse waves.

\begin{acknowledgments}
The fruitful discussions with José Renato Alcaras at the start of this study are gratefully acknowledged. The authors gratefully acknowledge the financial support provided by the São Paulo Research Foundation (FAPESP) and the National Council for Scientific and Technological Development (CNPq). NER was supported by the National Council for Scientific and Technological Development – CNPq (Grant No. 140549/2022-6); 
GN was supported by FAPESP grant 2023/07241-5; OMB thanks FAPESP by the grants 2018/22214-6, 2021/08325-2 and CNPq by the grant 307897/2018-4; A.S.M. acknowledges Brazil’s National Council for Scientific and Technological Development CNPq (Grant No. 0304972/2022-3) and the financial support by National Institute of Science and Technology in Innovative Research in Health Sciences from Nanotechnology to Artificial Intelligence (INCT PICS) CNPq, Grant No. 408417/2024-2. This support was essential for the successful completion of this research.
\end{acknowledgments}
\noindent\textbf{Data Availability}. The numerical simulations supporting this study—including all Python scripts and resulting graphs—are openly available on Zenodo at~\href{https://doi.org/10.5281/zenodo.22004404}{doi.org/10.5281/zenodo.22004404}.

\noindent \textbf{Conflict of interest declaration}.
We declare we have no competing interests.

\noindent \textbf{Author contributions}. N.E.R. conceived the study, developed the numerical
simulations, and wrote the original draft of the manuscript. O.M.B., G.N.,
and A.S.M. reviewed and edited the manuscript.

\appendix

\section{Sound-speed-driven contrast}
\label{sec:velocity}
\begin{figure*}[htb]
\includegraphics[width=1.0\textwidth]{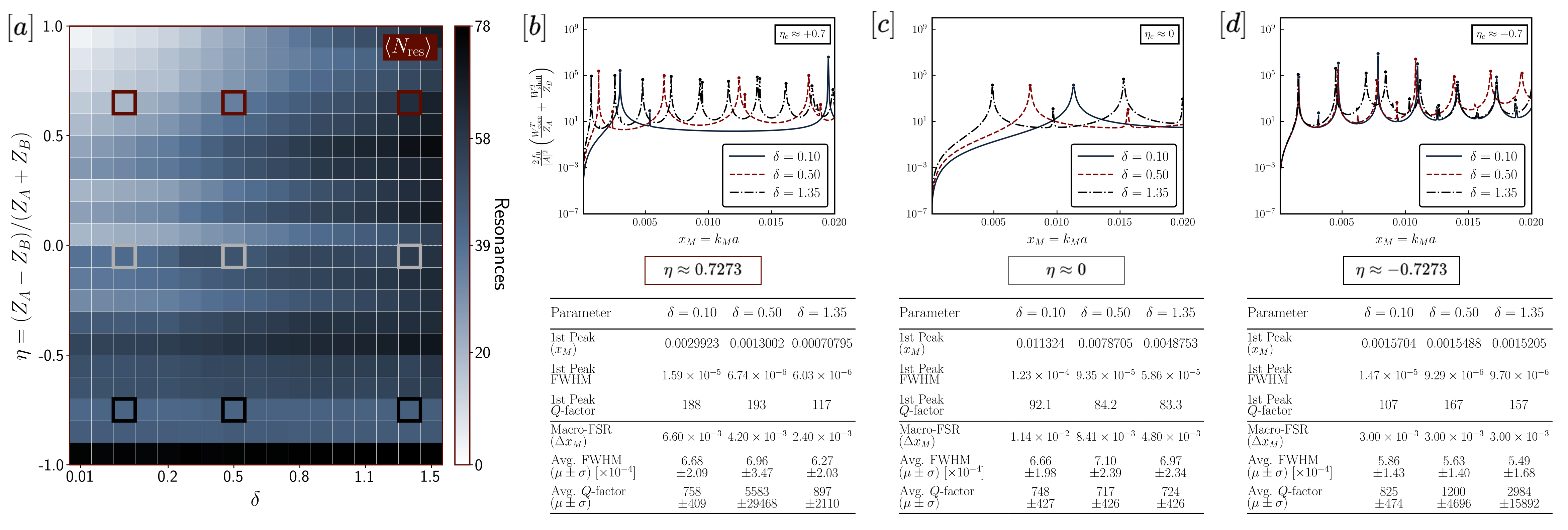}
\caption{State map and representative spectra for the sound-speed-driven (velocity) regime at fixed densities. [a] Average number of resolved resonances per bin, $\langle N_{\text{res}} \rangle$, as a function of the impedance-asymmetry parameter $\eta$ and the shell thickness $\delta$. Colored boxes mark the representative configurations examined in panels [b]--[d]. [b--d] Normalized internal acoustic energy as a function of the dimensionless size parameter $x_M = k_M a$, for three shell thicknesses ($\delta = 0.10$, $0.50$, $1.35$) at fixed $\eta \approx +0.7$ [b], $\eta \approx 0$ [c], and $\eta \approx -0.7$ [d]. The tables below each panel summarize, for every $\delta$, the position and FWHM of the first resonance peak, its $Q$-factor, the macroscopic free spectral range ($\Delta x_M$), and the mean $\pm$ standard deviation of the FWHM and $Q$-factor averaged over all resolved resonances.
\label{fig:speedboard}}
\end{figure*}

The complementary state maps presented in this Appendix provide a more detailed view of the resonant structure associated with the sound-speed-driven, mixed, and density-driven regimes. In these maps, $\langle N_{\text{res}} \rangle$ denotes the average number of resolved resonances per bin as a function of the impedance-asymmetry parameter $\eta$ and the relative shell thickness $\delta$. However, each $(\eta,\delta)$ bin contains several distinct $(\alpha,\beta)$ configurations, since $\eta=(\alpha-\beta)/(\alpha+\beta)$ does not uniquely determine the individual core and shell impedance contrasts. Consequently, the maps should not be interpreted as direct resonance-existence diagrams: resonant and non-resonant configurations may coexist within the same $(\eta,\delta)$ bin, even when $\langle N_{\text{res}} \rangle$ is nonzero or relatively large. Their purpose is therefore to characterize how the spectral structure is organized with impedance asymmetry and shell thickness within each contrast regime.

For this reason, the resonance status of a specific physical configuration must first be established in the $(\alpha,\beta)$ contrast map of Fig.~\ref{fig:max_resonances_contrast}. That map identifies which combinations of core and shell impedance contrasts are capable of supporting resonances. Only after the corresponding $(\alpha,\beta)$ point has been located does the $(\eta,\delta)$ map provide information about how the shell thickness modifies the resonant spectrum. In other words, $\delta$ can shift resonance positions, change their number, widths, and spacing, but it does not by itself make an otherwise non-resonant $(\alpha,\beta)$ configuration resonant. This distinction is particularly important in the forbidden quadrant $\alpha>1$, $\beta>1$, which occupies at least one quarter of the $(\alpha,\beta)$ plane considered here and remains strictly non-resonant for all values of $\delta$. The representative configurations selected in Figs.~\ref{fig:speedboard}[b]--[d] are therefore interpreted only after this $(\alpha,\beta)$ resonance condition has been taken into account, allowing the subsequent analysis to isolate how the spectral organization changes with $\eta$ and $\delta$.

For $\eta \approx +0.7$ (Fig.~\ref{fig:speedboard}[b]), the shell thickness strongly controls the resonance sequence. As $\delta$ increases, more peaks appear, and the spectra form richer multiplet structures, arising from different multipole orders $l$ that lie close enough in frequency to overlap and form the groups seen in [a]. Accordingly, the average $Q$-factor rises sharply from 758 at $\delta=0.10$ to 5583 at $\delta=0.50$, along with a narrower average linewidth and a smaller macro-FSR, in line with the denser, higher-quality resonance sequence. In this regime, $\eta>0$ corresponds to $Z_A>Z_B$, and the internal-field distribution indicates that a larger fraction of the acoustic energy is present within the lower-impedance shell, although the field remains distributed throughout the particle. Changes in $\delta$ therefore directly modify the region containing the largest share of the internal energy, which is consistent with the pronounced dependence of the resonance sequence on shell thickness.

Conversely, for $\eta \approx -0.7$ (Fig.~\ref{fig:speedboard}[d]), the spectra remain sparse and depend only weakly on $\delta$. The total peak count and multiplet composition stay nearly unchanged, while the resonance positions and linewidths vary only modestly. Still, the average $Q$-factor rises steadily with $\delta$ (825 $\to$ 1200 $\to$ 2984), but without the strong peak-count growth seen for $\eta>0$, showing that thicker shells sharpen the existing resonances rather than create new ones. In this case, $\eta<0$ corresponds to $Z_A<Z_B$, so that the lower-impedance region is now the core and a larger fraction of the internal field and energy is associated with this region. As a result, variations in shell thickness no longer act directly on the region where most of the energy is distributed, providing a consistent interpretation for the much weaker dependence of the spectral structure on $\delta$.

Meanwhile, the near-symmetric regime, $\eta\simeq0$ (Fig.~\ref{fig:speedboard}[c]), is dominated by singlets. Here, the average $Q$-factor stays essentially flat ($\sim$717--748) regardless of shell thickness, in contrast to the strong $\delta$-dependence observed for $\eta>0$. Since $Z_A$ and $Z_B$ become comparable in this limit, no similarly pronounced preference of the internal energy toward either the core or the shell is expected. The resulting spectral behavior therefore reinforces the role of the impedance hierarchy in determining how strongly the resonance structure responds to variations in shell geometry.

\section{Mixed contrast}
\label{sec:mixed}
\begin{figure*}[htb]
\includegraphics[width=1.0\textwidth]{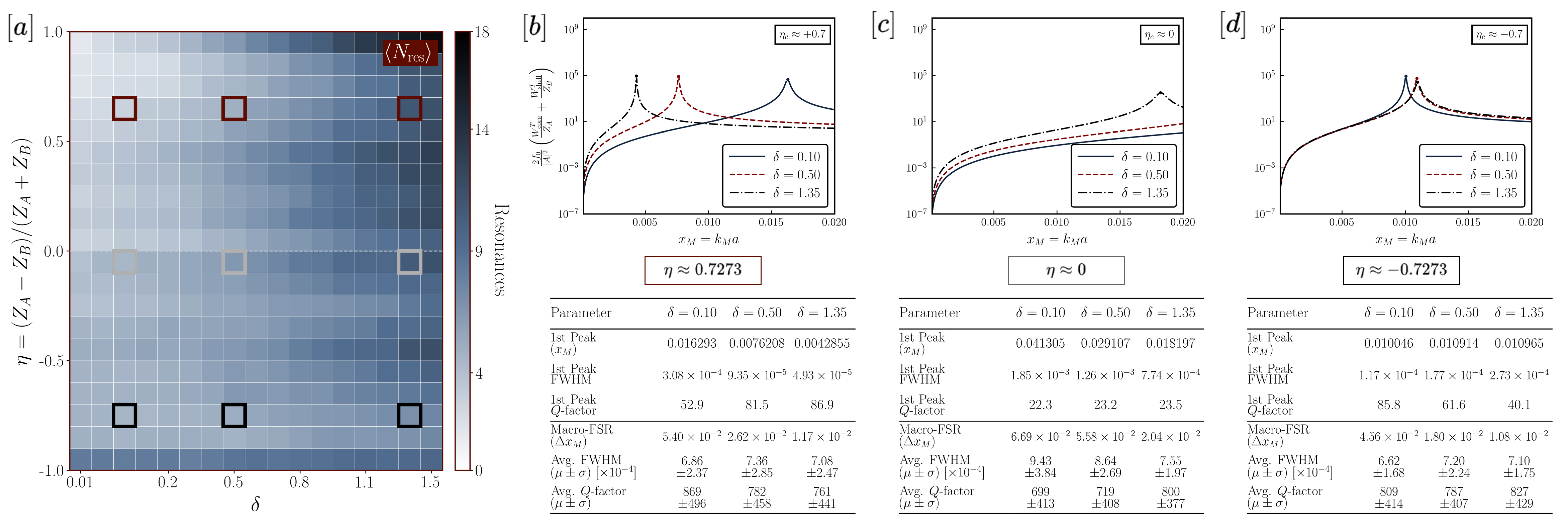}
\caption{\label{fig:mixed_board}
Mixed-case resonant response for the balanced contrast parametrization $f_{\text{mix}}=0.5$, in which the impedance variation is introduced equally through density and sound-speed contrasts. (a) Resonance-state map showing the average number of resolved resonances in each $(\delta,\eta)$ bin up to $x_M=1$. The selected configurations are examined through their energy spectra in (b)--(d). For $Z_M>Z_B>Z_A$, the resonance pattern is only weakly affected by $\delta$. In contrast, for $Z_M\gg Z_A<Z_B$, increasing the shell thickness shifts the resonances toward smaller $x_M$, making $\delta$ an effective tuning parameter.}
\end{figure*}

The mixed regime, in which the impedance contrast is introduced equally through density and sound-speed variations ($f_{\text{mix}}=0.5$), follows an overall organization similar to that of the sound-speed-driven case: the resonance-state map in Fig.~\ref{fig:mixed_board}[a] shows the same qualitative dependence on the sign of $\eta$, with more resonances appearing for $\eta>0$. However, the average resonance count is markedly lower across the entire map, indicating that mixing density and sound-speed contrasts substantially reduces the number of resolved peaks compared with the pure sound-speed-driven case.

This reduction becomes evident when comparing the individual spectra. For $\eta \approx +0.7$ (Fig.~\ref{fig:mixed_board}[b]), the peaks remain clearly distinguishable and shift with $\delta$, with the first peak moving from $x_M = 0.0163$ at $\delta=0.10$ to $x_M = 0.0043$ at $\delta=1.35$, and the average $Q$-factor staying comparatively stable ($869 \to 782 \to 761$). For $\eta \approx 0$ (Fig.~\ref{fig:mixed_board}[c]), a single dominant peak is still present, but its shift with $\delta$ is considerably weaker, moving from $x_M = 0.0413$ to $x_M = 0.0182$, while the average $Q$-factor increases only mildly ($699 \to 719 \to 800$). For $\eta \approx -0.7$ (Fig.~\ref{fig:mixed_board}[d]), the peaks essentially coincide regardless of $\delta$, with the first peak remaining between $x_M = 0.0100$ and $x_M = 0.0110$, and the average $Q$-factor varying only modestly ($809 \to 787 \to 827$).

\section{Density contrast}
\label{sec:density}
\begin{figure*}[htb]
\includegraphics[width=\textwidth, height=\textheight, keepaspectratio]{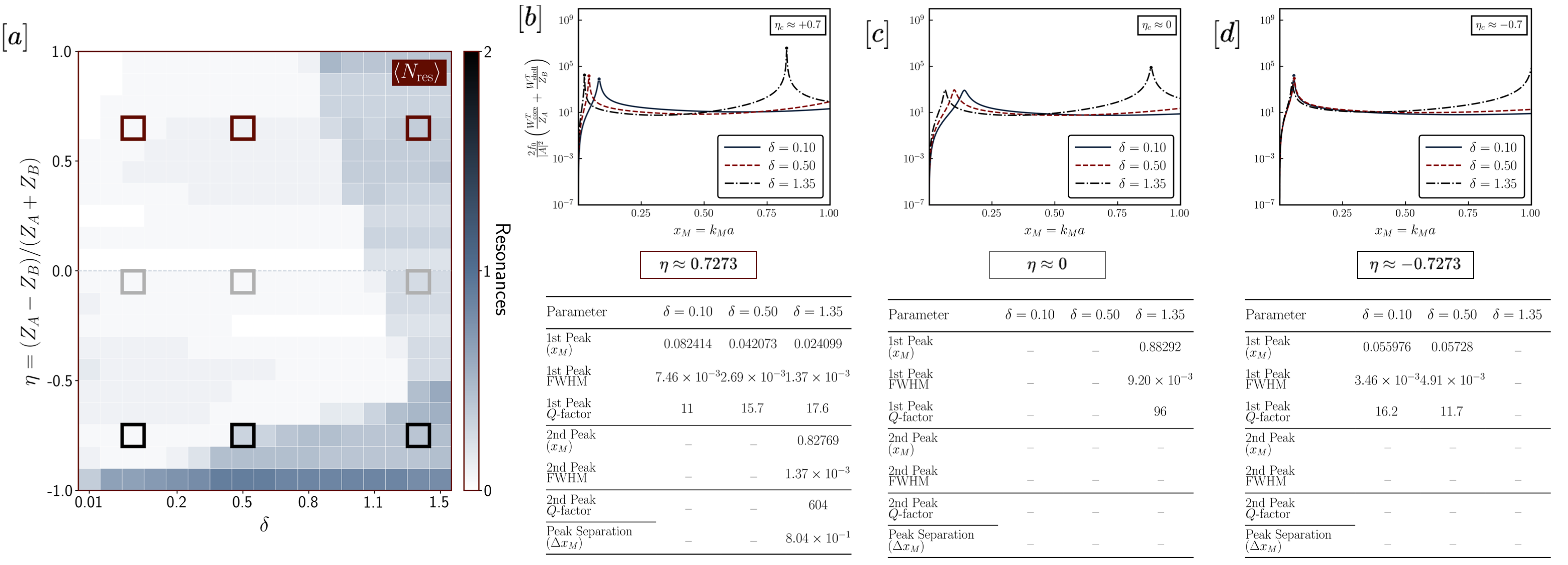}
    \caption{ \label{fig:density_board}Density-driven resonant response for fixed sound speeds. [a] State map showing the average number of resonances per $(\delta, \eta)$ bin up to $x_m = 1$. Peaks occur in a restricted region of relative impedance contrast ($\eta$), with weak dependence on the shell thickness ($\delta$). [b]–[d] Resonance spectra for selected $\eta$ and $\delta$ values. The profiles indicate predominantly core-localized resonances, showing that the shell does not act as an efficient tuning layer.}
\end{figure*}

At fixed sound speeds, the density-driven regime differs strongly from the mixed and sound-speed-driven cases. When the resonances are counted up to $x_M=1$, the active region in Fig.~\ref{fig:density_board}[a] becomes much smaller and contains only a few resonant configurations. Within this range, each selected configuration presents at most one or two resolved resonances, in clear contrast with the much denser spectra found in the other contrast mechanisms.

This first resonance is a purely low-frequency feature, occurring for well-defined, narrow ranges of $x_M$, and its position depends on both $\eta$ and $\delta$. For $\eta \approx +0.7$ (Fig.~\ref{fig:density_board}[b]), it shifts from $x_M=0.0824$ at $\delta=0.10$ to $x_M=0.0241$ at $\delta=1.35$, and a second, well-separated and much higher-$Q$ resonance appears only for the thickest shell. A similar shift is seen near impedance balance (Fig.~\ref{fig:density_board}[c]). For $\eta \approx -0.7$ (Fig.~\ref{fig:density_board}[d]), however, $\delta$ plays essentially no role: the first resonance barely moves between $\delta=0.10$ and $\delta=0.50$, and no resonance at all is resolved for $\delta=1.35$, since the shell in this configuration is not thick enough to sustain one. This shows that, unlike in the sound-speed-driven and mixed regimes, increasing the shell thickness does not guarantee the appearance of new resonant states in the density-driven case, and for negative $\eta$ it may even suppress the resonance entirely.

\bibliography{bibdatabase}

\end{document}